Short Paper

# Face Recognition for Motorcycle Engine Ignition with Messaging System

**Yolanda D. Austria**

Computer Engineering Department Adamson University, Philippines
yolanda.austria@adamson.edu.ph
(corresponding author)

**Luisito L. Lacatan**

Computer Engineering Department Adamson University, Philippines
luisito.lacatan@adamson.edu.ph

**John Gregory D. Funtera**

Computer Engineering Department Adamson University, Philippines
greg.funtera@gmail.com

**Shawn C. Garcia**

Computer Engineering Department Adamson University, Philippines
shawn.garcia0012@yahoo.com

**Jonet H. Montenegro**

Computer Engineering Department Adamson University, Philippines
jonet.montenegro@gmail.com

**Laymar T. Santelices**

Computer Engineering Department Adamson University, Philippines
laymarsantelices@yahoo.com

**Abstract**

*Purpose* – In this current world where technology is growing up day by day and scientific researchers are presenting new era of discoveries, the need for security is also increasing in all areas. At present, the vehicle usage is basic necessity for everyone. Simultaneously, protecting the vehicle against theft is also very important. Traditional vehicle security system depends on many sensors and cost is also high. When the vehicle is stolen, no more response or alternative could be available to help the owner of the vehicle to find it back. The main goal of this paper is to protect the vehicle from any unauthorized access, using fast, easy-to-use, clear, reliable and economical face recognition technique.

*Method* – An efficient automotive security system is implemented for anti-theft using an embedded system for starting the engine by the use of face recognition and integrated with Global Positioning System (GPS) and Global System for Mobile Communication (GSM). This proposed work is an attempt to design and develop a smart anti-theft system that uses Face recognition, GPS and GSM system to prevent theft and to determine the exact location of vehicle.

*Results* – The acceptance test of Face Recognition System for authentication of engine ignition acceptance test#1 shows that the system can recognize the faces of the registered user and ignite the engine. The USB webcam will capture the image of an unregistered/unauthorized user who tried to start the engine and send a text warning notification with an image to the user. The acceptance test of GPS and GSM for location and anti-theft notification acceptance test#1 shows that the GPS can receive the latitude and longitude values of its current location. The GSM can send the current location which had been received by the GPS as text message to user and also send an anti-theft notification if the motorcycle had been moved from its initial to approximately 5 m while engine is off.

*Conclusion* – This study concludes that in order to achieve the prevention of crashing of the system, the Face Recognition System for authentication of engine ignition acceptance test#1 should have an indicator to know if the program is executed successfully. Also, in order to achieve receiving the exact location of the motorcycle, the GPS and GSM for location and anti-theft notification acceptance test#1 should have an antenna in order for the GPS to receive the coordinates from the satellites much more accurate than that of without antenna. Lastly, in order to achieve the error prevention in receiving text messages of the GSM module, the Engine Ignition by Passcode and GSM acceptance test#1 should always delete received messages to clear the allotted memory storage for messages.

*Recommendations* – This study recommends further improvements in terms of security measures and integration of smart systems.

# INTRODUCTION

Philippines, as we know today, has evolved in terms of communication, transportation, entertainment, and other aspects that improved the lives of the Filipino people. The rapid advancement of these aspects is through the help of emerging technologies. Most people today rely on technology because of the advantages it brings which make our lives easier.

Security in today's world has also become more advanced because of technology. In preventing thefts for instance, various types of security systems have been developed. There are CCTVs (Closed-circuit Television) which can be found in most commercial establishments because of its high effectivity in preventing and solving crimes, burglar alarms used by commercial establishments which help prevent burglary thefts unauthorized access by setting off a loud alarm, button alarms which automatically alert the nearest police station that crime was attempted or is currently taking place, and many more. There are also different kinds of authentication that are used to increase security features in different kinds of devices such as fingerprint, retinal, iris, and face recognition. Among the types of security features mentioned, face recognition is one of the most sophisticated and secured.

The number of cases of vehicle that is being stolen in the Philippines is mostly on motorcycle vehicles based on the online news portal of TV5 (InterAksyon.com, 2015). The Highway Patrol Group (HPG) of the Philippine National Police (PNP) reported to the Senate that there were more motorcycles stolen in Metro Manila than cars in the first quarter of 2015 (InterAksyon.com, 2015). Having an authentication to the vehicle can increasingly prevent car thefts.

Motorcycle vehicle theft is one of the most common incidents of stealing in the country. The Philippine National Police has been registering a periodic increase in cases of stolen motor vehicles and motorcycles across the country, there are more cases of motorcycle vehicle theft compared with car theft incidents, which can be easily stolen when parked unattended. Having an authentication before starting the motorcycle can be used to increase its security but there are still instances that it is still stolen. The most common way of stealing a motorcycle is by lifting it off of the ground and loading it into a van (Aurelio, 2014). Through that method, the thefts can steal the motorcycle quickly and quietly with less chance of getting caught (Siler, 2012).

This means that an average of 4.4 vehicles were stolen in January, 4.6 in February, and 3.7 in March – or 4.2 vehicles stolen daily for the first three months of the year (Felipe, 2016). In 2015, the PNP recorded 9,201 carnapping incidents. Of these, 998 were stolen motor vehicles or four-wheeled vehicles like cars, wagons, SUVs and pick-ups while 8,203 are motorcycles. The news correspondent said, of the 9,201 incidents reported, 8,944 were validated as genuine carnapping incidents comprising of 968 motor vehicles and 7,976 motorbikes. In November, the PNP-HPG recorded an average of 3.18 4-wheeled vehicles stolen per day and 6.2 motorcycles being stolen per day. He said the PNP-HPG is recording all car napping incidents nationwide as reported and validated by Regional Highway Patrol Units (Dializon, 2016).

According to Singh, Sethi, Biswal, and Pattanayak (2015), security (especially theft security of vehicle in common parking places) has become a matter of concern. An efficient automotive security system is implemented for anti-theft using an embedded system integrated with Global Positioning System (GPS) and Global System for Mobile Communication (GSM). Moreover, to further improve the security of the vehicle, Sundari, Laxminarayana, and Laxmi (2012) stated that the use of vehicle is a must for everyone. At the same time, protection from theft is also very important. Prevention of vehicle theft can be done remotely by an authorized person. The location of the car can be found by using GPS and GSM controlled by FPGA.

GPS is a network of orbiting satellites that send precise details of their position in space back to earth which are obtained by GPS receivers. It was first used for military operations at the height of the Cold War. Ever since the early 1980s, however, the technology has been freely available to anyone with a GPS receiver. Airlines, shipping companies, trucking firms, and drivers everywhere use the GPS system to track vehicles, follow the best route to get them from A to B in the shortest possible time (MiTAC Digital Corporation, 2018).

According to Kamble (2012), the roots of Vehicle Tracking Systems lie in shipping industry. They required some sort of system to determine where each vehicle was at any given time and for how long it travelled. Initially, vehicle tracking systems developed for fleet management were passive tracking system. According to Verma and Bhatia (2013), GPS is one of the technologies that are used in a huge number of applications today. One of the applications is tracking your vehicle and keeps regular monitoring on them. This tracking system can inform you the location and route travelled by vehicle, and that information can be observed from any other remote location. It also includes the web application that provides you exact location of target. This system enables us to track target in any weather conditions. This system uses GPS and GSM technologies.

A GSM modem is used to send the position (latitude and longitude) of the vehicle from a remote place. The GPS modem will continuously give the data i.e. the latitude and longitude indicating the position of the vehicle (Biometric Solutions, 2016). This system designed for users in land construction and transport business, provides real-time information such as location, speed and expected arrival time of the user is moving vehicles in a concise and easy-to-read format. This system may also be useful for communication process among the two points.

Currently GPS vehicle tracking ensures their safety as travelling. This vehicle tracking system found in clients vehicles as a theft prevention and rescue device. Vehicle owner or Police follows the signal emitted by the tracking system to locate a robbed vehicle in parallel the stolen vehicle engine speed going to decreased and pushed to off. After switching of the engine, motor cannot restart without permission of password. This system installed for the four wheelers, Vehicle tracking usually used in navy operators for navy management functions, routing, send off, on board information and security. The applications include monitoring driving performance of a parent with a teen driver. Vehicle tracking systems accepted in consumer vehicles as a theft prevention and retrieval device. If the theft identified, the system sends the SMS to the vehicle owner. After that vehicle owner sends the SMS to the controller, issue the necessary signals to stop the motor (Biometric Solutions, 2016).

## OBJECTIVES

The general objective of this study is to develop a system with hardware and software components that would optimize the security of the motorcycle vehicles. In line with this, the project aims to achieve the following specific objectives:
- To install a face recognition system in the motorcycle vehicle for authentication of engine ignition.
- To implement GSM and GPS that will notify the exact location of the motorcycle during incidents of theft.
- To implement other means of engine ignition aside from face recognition. These are through GSM and through entering a passcode.

## METHODOLOGY

This chapter includes the research methodology of the study. It discusses the research strategy and approach in implementing the project by the researches. This section shows the different developmental phase in order to develop the design project. Furthermore, this also contains a several evaluation and tests to ensure the stability and reliability of the project.

The project was developed by analysing the requirements and by fully understanding the problems. The solution of the problems during the development was initially analysed and identified to use an appropriate components and applications. Once the needed information had been identified, implementation of the gathered requirement will take place. The project will undergo a testing phase to verify and reduce any possible errors that had been used during the implementation process. Lastly the information gathered from testing phases will be evaluated so that project will produce the expected results and perform its objectives.

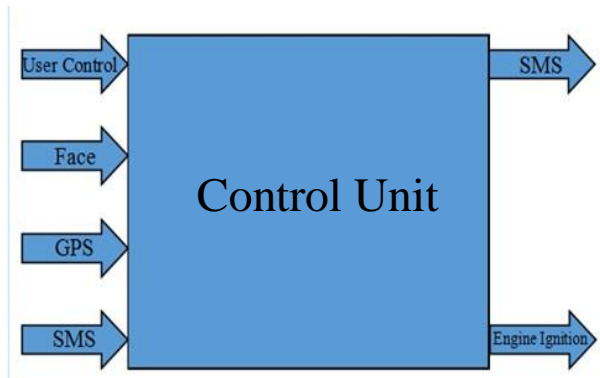

*Figure 1.* Level 0 block diagram of the Engine Ignition System for Motorcycles

Figure 1 illustrates the level 0 diagram of the Engine Ignition System for Motorcycle. It shows the inputs which is the user control, face of the person, SMS and GPS. The outputs will be the Engine Ignition System for Motorcycle and SMS. Level 0 diagram presents a single module block diagram with inputs and outputs identified.

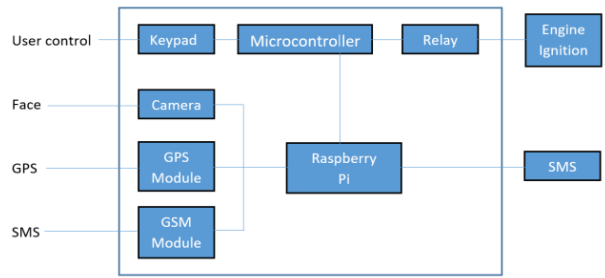

*Figure 2.* Level 1 diagram of the Engine Ignition System

Figure 2 illustrates the level 1 diagram of the Engine Ignition System. It shows the inputs which are the User Control, Face, GPS and SMS. The output of the Level 0 diagram will be the Engine Ignition for Motorcycle and SMS. Level 1 diagram provides system architecture with all modules and its interconnections.

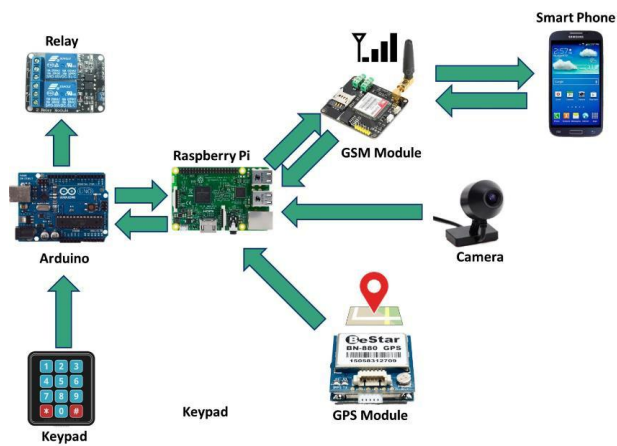

*Figure 3.* System Architecture Diagram

Figure 3 shows block diagram of the study illustrates the connection of all the devices such as Camera, Raspberry Pi, Relay, Arduino, Smart Phone, Keypad, GSM and GPS Module. The database of every device is centralized through the Raspberry Pi. Furthermore, it shows the illustration of the Project to visualize the connection between these several devices and their features. The Raspberry Pi is the central storage of all the information needed for Face Recognition System. The Smart Phone of the owner is used to send a keyword for the engine ignition aside from face recognition.

Figure 4 shows the flow chart of the system. There are three ways to start the engine. The first method is through face recognition. If the face is recognized, then the engine will start. Otherwise, the system will send a text message to the owner with the URL of the uploaded picture of the unregistered user. Second is by entering a passcode through keypad. If the passcode is correct, then the engine will ignite. The third method of starting the engine is through sending the commands via SMS. If the user sends a text to the system containing the keyword for starting the engine, then the engine will ignite. The user can also request the location of the motorcycle by just sending an SMS containing the keyword for requesting location. And if the engine of the motorcycle is off and the location is changed, the system will automatically send an SMS to the owner containing the location of the motorcycle and alarming the owner of a possible theft incident.

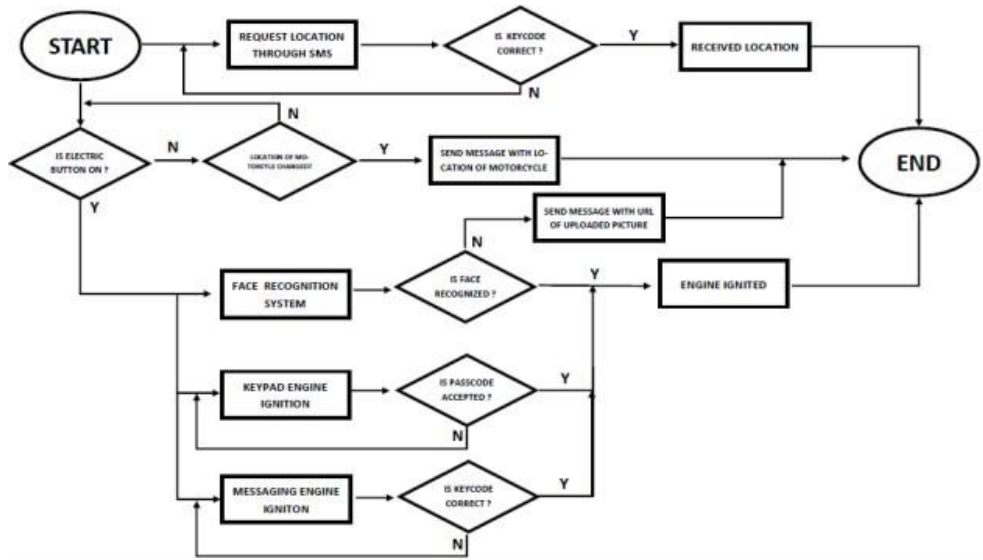

*Figure 4.* System flowchart of the project

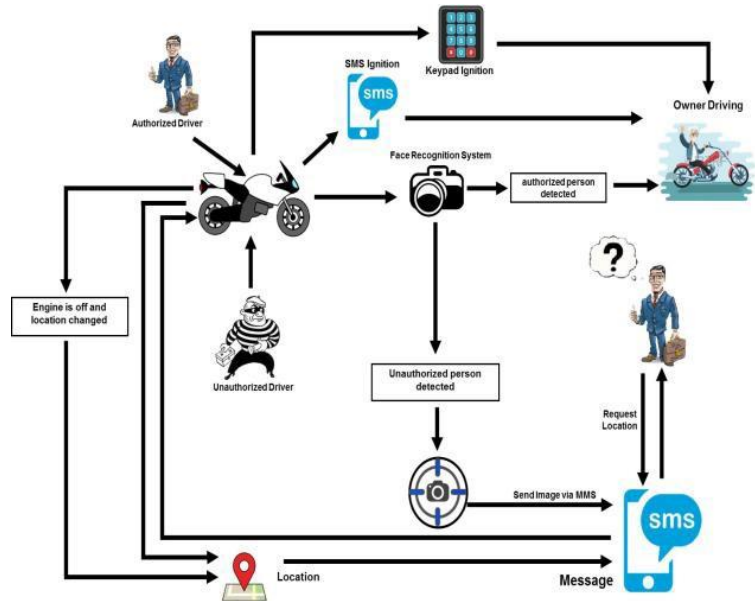
*Figure 5*. Workflow diagram

Figure 5 shows the workflow diagram of the project prototype. The authorized user can ignite the engine of the motorcycle by face recognition system, SMS ignition by sending a keycode, keypad ignition by entering the correct passcode. The user can also detect the location of the motorcycle by sending an SMS keycode and it will respond back by a text message with current location of the motorcycle. A SMS notification with an image will be received by the user if an unauthorized user attempts to ignite the engine of the motorcycle. Also if the location of vehicle has been changed while the engine is turned off it will also send an SMS notification of its current location to the user.

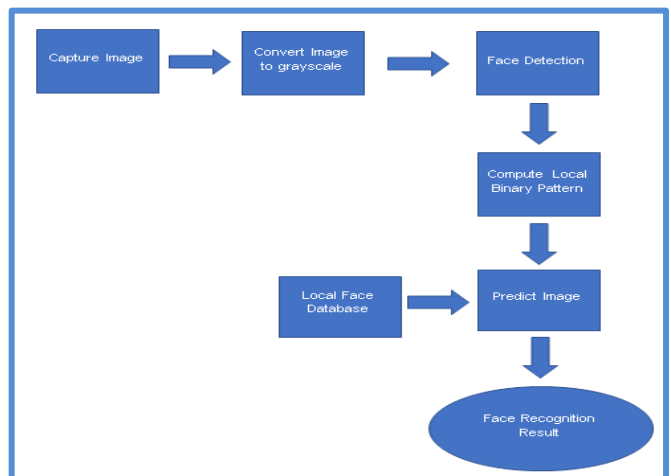
*Figure 6*. Face Recognition System Diagram

Figure 6 shows the diagram for the face recognition system. The capture image will be converted to grayscale for image processing. The system will identity the face within the image and applies the local binary pattern (LBP) for face recognition algorithm. The result of the algorithm will be used for comparing the images that are stored to the local database. It will finally give the decision whether the captured image is stored in to the local face database or not.

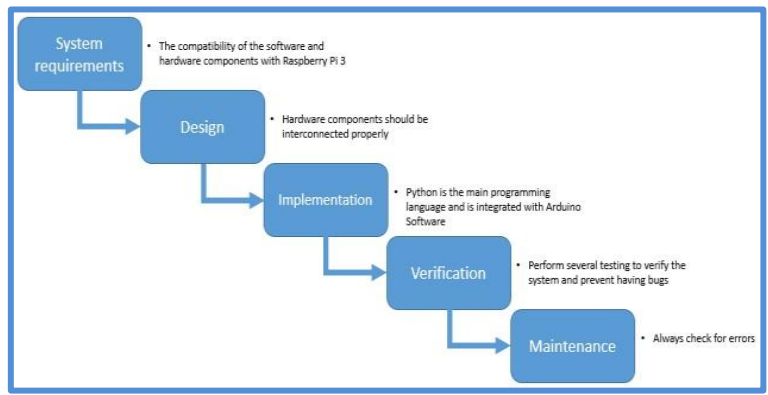

*Figure 7.* LBP Representation

Figure 7 shows the waterfall diagram of the system. Each hardware and software components should be compatible with Raspberry Pi 3 because it would be the main component of the system. Each component should be strategically placed in the control unit box and interconnected properly with Raspberry Pi 3. The main programming language will be used is Python and is integrated with Arduino Software. Perform several testing to ensure that the system is working properly and to avoid having bugs. Always check for errors for the maintenance of the system.

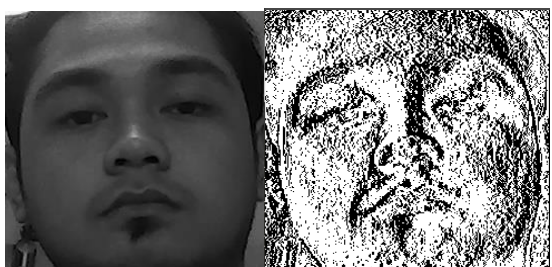

*Figure 8.* LBP Representation

Figure 8 shows the LBP representation of an image. An input was first converted into grayscale. For each pixel in the grayscale image, a neighbourhood is selected around the current pixel and calculate the LBP value for the pixel.

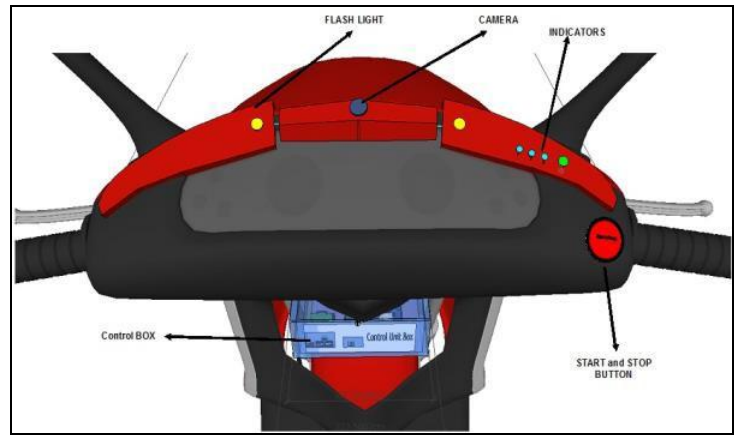
*Figure* 9. Physical view of the Vision System

Figure 9 illustrates the physical view of the vision system of the project. It shows the possible design which we can place the components and devices in the motorcycle. The camera is placed on the motorcycle dash board which projected to the rider and adjustable so he/she can angle the focus to him/her. Indicators are also embedded to notify the user the number of attempts and if the camera is activated. The control unit or the microcontrollers are protected on a box which a water and heat resistance, this place on the motorcycle that surely protected and hard to find.

## SUMMARY OF FINDINGS

This study summarizes the following findings:
- The acceptance test of Face Recognition System for authentication of engine ignition acceptance test#1 shows that the system can recognize the faces of the registered user and ignite the engine. The USB webcam will capture the image of an unregistered/unauthorized user who tried to start the engine and send a text warning notification with an image to the user.
- The acceptance test of GPS and GSM for location and anti-theft notification acceptance test#1 shows that the GPS can receive the latitude and longitude values of its current location. The GSM can send the current location which had been received by the GPS as text message to user and also send an anti-theft notification if the motorcycle had been moved from its initial to approximately 5 m while engine is off.

## CONCLUSIONS

This study concludes the following:
- In order to achieve the prevention of crashing of the system, the Face Recognition System for authentication of engine ignition acceptance test#1 should have an indicator to know if the program is executed successfully.

- In order to achieve receiving the exact location of the motorcycle, the GPS and GSM for location and anti-theft notification acceptance test#1 should have an antenna in order for the GPS to receive the coordinates from the satellites much more accurate than that of without antenna.
- In order to achieve the error prevention in receiving text messages of the GSM module, the Engine Ignition by Passcode and GSM acceptance test#1 should always delete received messages to clear the allotted memory storage for messages.

## RECOMMENDATIONS

This study recommends the following:
- To serve as a reference for future researchers in conducting a study about facial recognition system in further improving the security measures of a vehicle.
- To serve as a reference for motorcycle manufacturing companies in implementing better security measures for motorcycles. The motorcycle manufacturing companies can integrate a smart system that can send and receive SMS's, determine the location of the vehicle, and implement a keyless ignition on motorcycles.
- To lessen the instances of motorcycle robberies for this research will help educate owners by improving the security system of their vehicle through implementing passcode and SMS verification in starting the engine.